\documentclass[11pt]{article}
\usepackage{epsfig,amssymb,latexsym,amsmath,cite}

\newcommand{\be}{\begin{equation}}
\newcommand{\en}{\end{equation}}
\begin{document}

\thispagestyle{empty}

\baselineskip=15pt

\vspace*{-1cm}
\begin{center}
{\tiny  Topics in Mathematical Physics, General Relativity and
Cosmology, in Honor of Jerzy Pleba\'nski; H Garc\'{\i}a-Compe\'an
et al (Eds), World Scientific (Singapore, 2006) 373-385}
\end{center}

\begin{center}
{\Large \bf \sc On the Dirac--Infeld--Pleba\'nski delta function}
\end{center}

%\vskip1ex

\begin{center}
{\bf Oscar Rosas-Ortiz}
\vskip0.5ex
{\footnotesize Departamento de F\'{\i}sica, Cinvestav, AP 14-740,
07000 M\'exico~D~F, Mexico}
\end{center}

\vskip1ex
\begin{center}
\begin{minipage}{13cm}
{\footnotesize {\bf Abstract} The present work is a brief review
of the progressive search of improper $\delta$--functions which
are of interest in quantum mechanics and in the problem of motion
in General Relativity Theory. }
\end{minipage}
\end{center}

\begin{flushright}
\begin{minipage}{11.5cm}
{\em \footnotesize A great deal of my work was just playing with
equations and seeing what they give.}
\vspace*{-1.5ex}
\begin{flushright}
{\footnotesize P.A.M. Dirac}
\end{flushright}

%\vskip1ex

{\em \footnotesize Every one who works, no matter how briefly or
superficially, in complex relativity will find himself
acknowledging Jerzy's work.}
\vspace*{-1.5ex}
\begin{flushright}
{\footnotesize M.P. Ryan}
\end{flushright}
\end{minipage}
\end{flushright}

%----------------------------------------------------------->introduction

\section{Introduction}

The advent of quantum mechanics opened a new domain of concepts,
including generalized functions. Introduced as an {\it improper
function\/} by P.A.M. Dirac in 1926\cite{Dir26} (see also the
Dirac's book\cite{Dir58}), the delta function had been used in
physics for quite time before the formal work of L. Schwartz,
published in 1950\cite{Sch50}. The mathematical foundations of
generalized functions, however, appear to have first been
formulated in 1936 by S.L. Sobolev in his studies on the Cauchy
problem for hyperbolic equations\cite{Sob36} (see also the works
of J. Hadamard\cite{Had23} and M. Riesz\cite{Rie49}). At the
present time, the distribution theory has advanced substantially
and has found a number of applications in physics and mathematics.
Indeed, the use of generalized functions leads to remarkable
simplifications in the problems that one usually faces in
contemporary physics.

One of important applications of the Dirac's delta, out of the
quantum area, occurred in general relativity. In 1927, A. Einstein
and J. Grommer reported the first solution of the problem of
motion; in the procedure, they used the delta function to
represent matter in the field equations\cite{Ein27}. Thus, the
simplification of the problem was done at the cost of introducing
singular structures. Primary refinements were done in 1938 by
Einstein, L. Infeld, and B. Hoffman\cite{Ein38}. Some years later,
in their 1960 book {\it Motion and Relativity\/}, Infeld and J.
Pleba\'nski discussed a number of deeper improvements to the
Einstein, Infeld, and Hoffman approach\cite{Inf60}. In particular,
they modified the definition of the delta function in order to
properly manage singularities in general relativity.

In this paper we examine the origin of generalized --improper--
functions in quantum and general relativity theories. Although
history leads to a better understanding of the modern physics, it
is not my interest to cover the entire development but rather to
fix attention on some specific points. The paper is organized as
follows: Section~2 overviews the origin of the delta function in
quantum mechanics. Section~3 deals with the aspects of the motion
problem in general relativity (GR) connected with the
modifications developed by Infeld and Pleba\'nski on the Dirac's
delta function. Section~4 is devoted to some other ``delta
objects'' appearing quite recently in the literature and to the
concluding remarks.

%----------------------------------------------------------->ch1

\section{The Dirac's delta function}

One of the most interesting developments of quantum mechanics
concerns the concept of commutativity and starts in 1925, with an
idea of W. Heisenberg: {\em one ought to ignore the problem of
electron orbits inside the atom, and treat the frequencies and
amplitudes associated with the line intensities as perfectly good
substitutes. In any case, these magnitudes could be observed
directly}\cite{Hei71}. Indeed, {\em it is necessary to bear in
mind that in quantum theory it has not been possible to associate
the electron with a point in space, considered as a function of
time, by means of observable quantities. However, even in quantum
theory it is possible to ascribe to an electron the emission
radiation}\cite{Hei25} (see Van der Waerden\cite{Wae67} p 263). He
was certain that {\em no concept enter a theory which has not been
experimentally verified at least to the same degree of accuracy as
the experiments to be explained by the theory}\cite{Hei30}.
Thereby, Heisenberg concluded that the physical variables should
be represented by specific arrays of numbers (matrices). A
conclusion which, in turn, led him to an apparently unexpected
result: {\em Whereas in classical theory $x(t)y(t)$ is always
equal to $y(t)x(t)$, this is not necessarily the case in quantum
theory}\cite{Hei25} (see Van der Waerden\cite{Wae67} p 266). It
was almost inconceivable that the product of physical quantities
could depend on the multiplication order.

Thus, in exchange for the classical notion of position and
momentum in atoms, Heisenberg introduced the concept of {\it
observables\/} (measurable experimental magnitudes) and remarked
on their non-commutation properties. After the approval of W.
Pauli, Heisenberg published his results in the paper {\it Quantum
theoretical reinterpretation on kinematic and mechanical
relations\/} ({\it \"Uber quantentheoretische Undeutung
Kinematischen und mechanishen Beziehunge\/})\cite{Hei25} and gave
a copy to M. Born. When Born read the paper he noticed that the
Heisenberg's symbolic multiplication was the matrix algebra. Later
on, Jordan, Heisenberg, and Born published the {\it three men's\/}
paper {\it On Quantum Mechanics\/} ({\it Zur
Quantenmechanik\/})\cite{trio} in which they reported a matrix
formulation of the new theory (See also Van der
Waerden\cite{Wae67}, Jammer\cite{Jam89},
Duck--Sudarshan\cite{Duc00}, and Mehra\cite{Meh01a,Meh01b}).

On 28 July 1925, during a stay in Cambridge with R.H. Fowler,
Heisenberg delivered the talk ``Term Zoology and Zeeman Botany''
before the {\it Kapitza Club\/}. The subject dealt with the
anomalous Zeeman effect and the enormous difficulties to build
atomic spectroscopy up by means of {\it ad hoc\/} rules, a
remarkable topic for somebody who had solved the quantum puzzle
recently (see Mehra\cite{Meh01b} Ch 19.10). A month later,
Heisenberg sent the proof-sheets of his paper to Fowler who, in
turn, gave it to his research student, Dirac. After reading the
paper, Dirac pondered it for two weeks and noticed that
Heisenberg's idea had provided the key to the `whole
mistery'\cite{Meh01b}. In his own words: {\em non-commutation was
really the dominant characteristic of Heisenberg's new
theory}\cite{Dir71}. Dirac concluded that quantum mechanics could
be inferred from the Hamilton's form of classical dynamics by
considering new `canonical variables' obeying a non--commutative
`quantum algebra'. The results were published by Dirac between
1925 and 1927. One of his papers, {\em The physical interpretation
of the quantum dynamics}\cite{Dir26}, was decisive in the
formalization of the new theory. There, an arbitrary function of
the position and momentum is shown to be smeared over the entire
momentum space if the position is infinitely sharp (the
uncertainty principle!). The main point of the formulation was a
{\it transformation theory\/} which required the introduction of
the {\it improper function\/} $\delta (\zeta)$.

Dirac reconsidered the Heisenberg's idea of observables: {\em When
we make an observation we measure some dynamical variable\ldots
the result of such a measurement must always be a real
number\ldots so we should expect a real dynamical variable\ldots
Not every real dynamical variable can be measured, however. A
further restriction is needed} (see Dirac\cite{Dir58} p 34-35).
Then, he formalized the concept by defining an {\it observable\/}
as a real dynamical variable whose eigenstates form a complete
set, and stated that, at least theoretically, every observable can
be measured. If the eigenvalues of the observable $\zeta$ consist
of all numbers in a certain range, then the arbitrary eigenkets
$\vert X \rangle$ and $\vert Y \rangle$ of $\zeta$ may be
expressible as the integrals

\begin{equation}
\vert X \rangle = \int \vert \zeta'x \rangle d\zeta', \quad \vert
Y \rangle = \int \vert \zeta''y \rangle d\zeta''
\label{ket}
\end{equation}
$\vert \zeta' \rangle$ and $\vert \zeta'' \rangle$ being eigenkets
of $\zeta$ belonging to the eigenvalues $\zeta'$ and $\zeta''$
respectively, $x$ and $y$ labelling the two integrands, and the
range of integration being the range of eigenvalues. We say that
$\vert X \rangle$ and $\vert Y \rangle$ are in the representation
of the basic bras $\vert \zeta \rangle$ (a similar definition is
true for discrete eigenvalues). By considering the product
$\langle X \vert Y \rangle$, we take the single integral

\begin{equation}
\int \langle \zeta'x \vert \zeta''y \rangle d\zeta''.
\label{product}
\end{equation}
The integrand in (\ref{product}) vanishes over the whole range of
integration except at the point $\zeta' = \zeta''$. Following
Dirac's formulation (see Dirac\cite{Dir58} Ch 10), as in general
$\langle X \vert Y \rangle$ does not vanish, so in general
$\langle \zeta'x \vert \zeta''y \rangle$ must be infinitely great
in such a way as to make (\ref{product}) non-vanishing and finite.
To get a precise notation for dealing with these infinite objects,
Dirac introduced the quantity $\delta (\zeta)$, depending on the
parameter $\zeta$ and fulfilling the conditions

\begin{equation}
\left\{
\begin{array}{l}
\delta(\zeta) = 0 \quad \mbox{\rm for} \quad \zeta \neq 0\\[12pt]
\int \limits_{-\infty}^{\infty} \delta (\zeta)\, d\zeta=1.
\label{ddelta}
\end{array}
\right.
\end{equation}
And, a more general expression:
\begin{equation}
\int \limits_{-\infty}^{\infty} f(\zeta) \, \delta (\zeta) \,
d\zeta = f(0).
\label{ddelta2}
\end{equation}
The range of integration in (\ref{ddelta}) and (\ref{ddelta2})
does not need to be from $-\infty$ to $\infty$, but may be over
any domain $\Omega$ surrounding the critical point at which the
$\delta$ function does not vanish. Dirac acknowledged there is
something unusual about the delta function and decided to call it
an `improper function'. The following expressions are essentially
rules of manipulation for $\delta$ functions

\begin{equation}
\begin{array}{c}
\int \limits_{\Omega} f(\zeta) \, \zeta \, \delta(\zeta) \, d\zeta
=0;\\[12pt]
 \int \limits_{\Omega} f(\zeta) \, \frac{d^{(n)}
\delta(\zeta)}{d \zeta^{(n)}} \, d\zeta = \left. (-1)^n
\frac{d^{(n)} f(\zeta)}{d\zeta^{(n)}} \right|_{\zeta =0}.
\end{array}
\label{dalg}
\end{equation}
The first of equations (\ref{dalg}) means that $\zeta
\delta(\zeta)$, as a factor in an integrand, is equivalent to
zero. The second one is easy to verify by using integration by
parts $n$ times, and means that $\delta(\zeta)$ can be formally
differentiated as many times as one wishes. There are diverse ways
to face the $\delta$--function. For example, it appears whenever
one differentiates a discontinuous function like the Heavside
one\footnote{Indeed, $\Theta(\zeta)$ is also an improper function.
Hence, the derivative $d\Theta/d\zeta$ should be understood as in
the context of equation (\ref{dalg}).}

\begin{equation}
\Theta(\zeta) = \left\{
\begin{array}{l}
0 \quad \zeta < 0\\[4pt]
\frac12 \quad \zeta=0\\[4pt]
1 \quad \zeta > 0,
\end{array}
\right.
\label{theta}
\end{equation}
for which one gets $d\Theta(\zeta)/d\zeta = \delta(\zeta)$. Of
special importance is the {\it Fourier representation\/} of
$\delta(\zeta)$. It is obtained trough the eigenfunctions of the
operator $id/d\zeta$, that is $(2\pi)^{-1/2}e^{-ik\zeta}$,
henceforth

\begin{equation}
\delta (\zeta' - \zeta'') = \frac{1}{2\pi} \int \limits_\Omega
e^{ik(\zeta' - \zeta'')} dk.
\label{fourier}
\end{equation}
Thus, the $\delta$--function is just a shorthand notation for
limiting process which simplifies calculations. In general, one
can take a class of functions $\delta(\varepsilon, \zeta)$, such
that

\begin{equation}
\lim_{\varepsilon \rightarrow 0} \int \limits_{\Omega} f(\zeta) \,
\delta (\varepsilon, \zeta) \,d\zeta= f(0).
\label{lim}
\end{equation}
In practice, one uses the following mathematical scheme: all
calculations have to be performed not on $\delta(\zeta)$ but on
$\delta (\varepsilon, \zeta)$. The limiting procedure $\varepsilon
\rightarrow 0$ has to be made in the very last result. Two
plausible models are the sequence of Gaussian distribution
functions

\begin{equation}
\delta (\varepsilon, \zeta) = \frac{1}{\varepsilon \sqrt{2\pi}}
e^{-\zeta^2/2\varepsilon^2}
\label{gauss}
\end{equation}
and the (simplest) set of square well potentials
\begin{equation}
V(v, \zeta) = \left\{
\begin{array}{rl}
-v & \quad \vert \zeta \vert \leq \varepsilon\\[4pt]
0 & \quad \vert \zeta \vert > \varepsilon
\end{array}
\right.
\label{square}
\end{equation}
with $v=-1/(2 \,\varepsilon)$; $\varepsilon > 0$. Finally, the
integrand of equation (\ref{product}) will now be written

\begin{equation}
\langle \zeta' \vert \zeta'' \rangle = \delta (\zeta' - \zeta'')
\end{equation}
were we have dropped the labels $x$ and $y$. If we are interested
in two different representations for the same dynamical system
$P$, the quantities $\langle \eta\vert \zeta \rangle$ are called
the {\it transformation functions\/} from the representation $\{
\vert \eta \rangle\}$ into $\{ \vert \zeta \rangle \}$. The ket
$\vert P \rangle$ will now have the two {\it representatives\/}
$\langle \eta \vert P \rangle$ and $\langle \zeta \vert P
\rangle$, defining the corresponding transformation equations (see
Dirac\cite{Dir58}, p 75). According to Dirac, the transformation
functions are example of probability amplitudes. Thus, the
statistical interpretation of Born is also applicable in the
Dirac's transformation theory.

Nowadays, the formulae (\ref{ddelta})-(\ref{lim}) are easy to
analyze but in Heisenberg-Dirac times neither matrix formalism nor
improper functions were popular among theoreticians. As we have
said, Heisenberg was advised by Born on the connection between
matrix algebra and his non-commutative operations. In
contradistinction, the Dirac's training as engineer seems to be
fundamental; in his own words: {\em all electrical engineers are
familiar with the idea of a pulse, and the $\delta$--function is
just a way of expressing a pulse mathematically} (see
Jammer\cite{Jam89} p 316). However, although the $\delta$--{\it
function\/} is attributed to Dirac, it had been introduced in
physics much earlier. Prior to its appearance in quantum
mechanics, it was used by Hertz in statistical mechanics in
connection with the concept of temperature (see Jammer\cite{Jam89}
pp 317). Its occurrence in pure mathematics was noticed in 1815 by
A.L. Cauchy whose derivation of the Fourier--integral theorem is
based on the modern $\delta$--function\cite{Cauchy} (the
derivation is reproduced in Van der Pol and Bremmer\cite{pol55} Ch
8). Also in 1815, S.D. Poisson worked on the Fourier--integral
theorem and followed a similar procedure as that of
Cauchy\cite{Poisson}. In 1882, G.R. Kirchhoff used the Green's
theorem in the study of Huygen's principle. Kirchhoff too was
acquainted with the improper function delta, which he denoted by
$F$: {\em As to the function $F$ we assume that it vanishes for
all finite positive and negative values of its argument, but that
it is positive for such values when infinitely small and in such a
way that

\[
\int F(\zeta) d\zeta =1
\]
where the integration extends from a finite negative to a finite
positive limit}\cite{Kir82}. Kirchhoff remarked on the fact that
$2\delta(\mu^{-1}/\sqrt{2}, \zeta)$ approximates $F$ for very
large $\mu$ (see eq. (\ref{gauss})). In 1891, influenced by the
works of Cauchy and Poisson, H. Hermite proposed the
integral\cite{hermite}

\[
\int_{\alpha}^{\beta} \frac{2i\lambda}{(t-\theta)^2 + \lambda^2}dt
\]
and analyzed its limit $\lambda \rightarrow 0$ for small values of
$\theta$. It is easy to see that by taking $\theta=0$, provided
$\alpha$ and $\beta$ lie at different sides of $t=0$, the above
integral defines the class of functions $\delta(\lambda, t)$
converging to $\delta(t)$ in the sense of (\ref{lim}). Later on,
the delta function is put forward in the work of O.
Heavside\cite{heavside1} (see equation (\ref{theta}) and the
Heavside books\cite{heavside2}).

Apart of the previously quoted antecedents, the intuitive Dirac's
procedure opened an intriguing problem in pure mathematics. Some
criticism was presented, e.g. von Neumann wrote: the Dirac's {\em
method adheres to the fiction that each self--adjoint operator can
be put in diagonal form. In the case of those operators for which
this is not actually the case, this requires the introduction of
`improper' functions with self--contradictory properties. The
insertion of such mathematical `fiction' is frequently necessary
in Dirac's approach... It should be emphasized that the correct
structure need not consist in a mathematical refinement and
explanation of the Dirac method, but rather that it requires a
procedure differing from the very beginning, namely, the relevance
on the Hilbert theory of operators}\cite{von55}. The first
attempts to mathematically formalize the definition of the
$\delta$--function were done in 1926-1927 by Hilbert and published
later by Hilbert, von Neumann and Nordhein under the title {\it
The foundations of quantum mechanics\/} ({\it \"Uber die
Grundlagen der Quantenmechanik\/}\cite{Hil27}.) The challenge was
finally faced by Schwartz in the context of his theory of
distributions\cite{Sch50} (see also Sobolev\cite{Sob36}). Thus,
the ill--defined $\delta$--function and its derivatives were
replaced by well--defined linear functionals (distributions) which
have always other distributions as derivatives on the test
functions space.

The Dirac's function plays an alternative role in quantum
mechanics: it is an exactly solvable potential enjoying many
useful applications\cite{Flu74}. As a physical model, it has been
used to represent localized matter distributions or potentials
whose energy scale is high and whose spatial extension is smaller
than other relevant scales of the problem. Arrays of
$\delta$--function potentials have been used to illustrate Bloch's
theorem in solid state physics (the Dirac's comb) and also in
optics where wave propagation in a periodic medium resembles the
dynamics of an electron in a crystal lattice. The bound state
problem in one dimension (for potentials involving either
attractive or repulsive delta terms) has an exact implicit
solution whenever the eigenvalue problem without deltas can be
solved exactly\cite{Eps59}. In two and higher dimensions it
provides a pedagogical introduction to the techniques of
regularization in quantum field theory\cite{Tho79} (in one
dimension, the quantum system needs no regularization.) It has
been also studied in the context of supersymmetric quantum
mechanics\cite{Boy88,Dia99,Gol94,Neg01,Dut01} where the susy
partner of the attractive $\delta$--function is the purely
repulsive $\delta$--function\cite{Boy88,Dia99,Neg01}. Similar
results are obtained for potentials made up of additive
$\delta$--function terms\cite{Gol94,Dut01}.

%----------------------------------------------------------->ch

\section{The Infeld-Pleba\'nski's delta function}

In all descriptions of nature we use two alternative concepts:
{\it field\/} and {\it matter\/}. Matter is composed of particles
and the field is created by moving particles. The picture is
simple at the cost of having singular fields. Furthermore, what
can one say about the motion of the sources?

In Newtonian gravitational theory the concept of gravitational
field is reduced to the action at a distance. However, according
to relativity, no linear field theory can determine the motion of
its sources because no action can be propagated with a speed
greater than the speed of light. Hence, one must add the motion
equations to the field equations. The statement is no longer true
in nonlinear field theories as Einstein and Grommer have shown in
their paper of 1927, {\it General Relativity Theory and Laws of
Motion\/} ({\it Allgemeine Relativit\"atstheorie und
Bewegungsgesetz\/})\cite{Ein27}. They obtained an unexpected
conclusion: the equations of motion for a test particle are but a
consequence of the field equations! The Einstein and Grommer paper
opened as well more general problems in GRT. One of them was how
to find whether the equations of motion of two particles can also
be deduced from the field equations: a challenge which remained
open till 1936, when L. Infeld arrived in Princeton to begin a
collaboration with Einstein.

Before his presence in Princeton, Infeld was working with M. Born
in Cambridge. They faced the problem of modify Maxwell's
electrodynamics so that the self energy of the point charge is
finite. Their results are nowadays known as the Born-Infeld
electrodynamics. As Einstein rejected the idea from the very
beginning and Born and Infeld did not succeed in their attempts to
reconcile it with quantum theory, Born `warmly recommended' Infeld
to Einstein (see Born\cite{Bor71} p 121) and, in fact, Infeld
became Einstein's collaborator and assistant. Later on Einstein
wrote to Born: {\em We} [Infeld and Einstein] {\em have done a
very fine thing together. Problem of astronomical movement with
treatment of celestial bodies as singularities of the field}
(Born\cite{Bor71} Lett. 71, p 130). The `fine thing' concerned a
fundamental simplification of the foundations of GR.

The Infeld--Einstein collaboration on the problem of motion was
more persistent than any problem Infeld had tackled before: {\em
For three years I worked on this problem whose only practical
application that I know of is the analysis of the motion of double
stars by methods giving deeper insight than the old Newtonian
mechanics. For three years I have been bothering with double
stars}\cite{Quest}. In principle, the movement of sources is
determined by the geodesic lines of the space-time world; the
metrics of which satisfy the Einstein's field equations. The point
of departure was the Einstein's assertion that the first part of
this assumption is redundant; it follows from the field equations
by going to the limit of infinitely thin, mass-covered world
lines, on which the field becomes singular (see Born\cite{Bor71} p
131, and Infeld\cite{Inf57b}). Infeld remarks {\em the
calculations were so troublesome that we decided to leave on
reference at the Institute of Advanced Study in Princeton a whole
manuscript of calculations for other to use}\cite{Inf60}.

Such a quantity of work was finally rewarded. In 1938 Einstein,
Infeld and Hofffman (EIH) published the paper {\em The
Gravitational Equations and the Problem of Motion}\cite{Ein38}, in
which the two--body problem was solved for the first time.
However, as the relativity non--linear field equations are too
cumbersome to be solved exactly, approximation methods were
required. The basic idea behind the EIH method is to take into
account that for a function depending on coordinates and time,
developed in a power series in the parameter $c^{-1}$, the time
derivatives will be of a higher order than the space derivatives.
In general, by using singularities to represent matter, the method
{\em consisted in forming certain two--dimensional surface
integrals over surfaces enclosing these singularities. The field
equations prescribed the laws by which the surfaces enclosing the
singularities, and hence these singularities moved}\cite{Inf57b}.

The Einstein--Infeld collaboration continued some more years
bringing a progress in the problem of motion whose {\em final
solution will never influence our daily lives and will never have
any technical application. It is a purely abstract problem}. An
even more skeptical thought of Infeld was {\em I do not believe
that there are more than ten people in the world who have studied
our papers on the problem of motion}\cite{Quest}. As it seems
nowadays, Infeld had underestimated the importance of their own
results.

Not long after the EIH work was successfully completed; Infeld,
then at the University of Toronto, published a paper with Wallace
in which the EIH approach is applied to the problem of motion in
electrodynamics\cite{Inf40}. Ten years later, in 1951, Infeld and
Scheidegger worked on the problem of gravitational radiation
reaction in the EIH formalism\footnote{Their paper {\em Radiation
and Gravitational Equations of Motion}\cite{Inf51} gave rise to a
considerable flow of discussion concerning the radiation reaction
problem in GR, a subject in which Infeld was involved as soon as
he arrived in Princeton. See the interesting Kennefick's
paper\cite{Ken97} for details.}. That same year Infeld left
Canada\cite{Quest,Ken97,Infe,Sta} and returned to Poland to join
the Institute of Theoretical Physics in Warsaw University. Once in
Warsaw, his work attracted the attention of an amount of brilliant
graduate students and researchers all interested in gravitational
wave theory. Bia\l ynicki--Birula, Suffczy\'nski, Trautman, Werle,
Pleba\'nski and Kr\'olikowski --the last two, students of
Rubinowicz-- are some of the names of the Infeld's group.

Almost fifteen years had elapsed since the EIH paper was published
and the problem of motion attracted again the interest of Infeld.
He worked on the subject with his group for about six more years
to collect finally all their results in the book {\em Motion and
Relativity}, written together with Jerzy Pleba\'nski in
1960\cite{Inf60}. The collaboration was profitable for young
Pleba\'nski\footnote{Pleba\'nski was almost 23 when Infeld arrived
in Poland in 1950. Not long after they began in a close
collaboration which strongly influenced the Pleba\'nski's first
research contributions. Indeed, it was in 1957 that Infeld invited
Pleba\'nski to write their monograph.}; Infeld introduced him to
non--linear electrodynamics (see for instance\cite{Inf54}),
unitary operators\cite{Inf55} and to the problem of motion in GR
among other topics. The elaboration of {\em Motion and Relativity}
took almost four years of discussion on the contents and typeset.
Both Infeld and Pleba\'nski, finished the first chapter and
appendices\footnote{The main results of their appendix~1 ``{\sc
The $\delta$ function}'', were published in 1956 in the paper {\em
On modified Dirac $\delta$--functions}\cite{Inf56a}. Later
improvements were published in 1957\cite{Inf57}.} when, in 1957,
Pleba\'nski received a Rockefeller Fellowship to go to the United
States. Before leaving, Pleba\'nski prepared a sketch in Polish of
the rest of the manuscript, except the last chapter which was
added by Infeld lone\cite{Inf60,Ken97} (see also
Garc\'{\i}a-Compe\'an et al\cite{pref} in this volume).

The birth of the Infeld--Pleba\'nski delta functions occurred at
this phase of the research. In some sense, it followed the early
Einstein's ideas: the energy--momentum tensor $T_{\alpha \beta}$
in the field equations introduced indeed an excess of information
mixing the physics and geometry. The use of $G_{\alpha \beta} =
R_{\alpha \beta} -\frac12 g_{\alpha
\beta} = -8\pi \, T_{\alpha \beta}$, with $T_{\alpha \beta}$
proportional to the Dirac's function $\delta_{(3)}$ (the presence
of matter), permits to skip the redundance, reducing the geometry
to the singular solution of $G_{\alpha
\beta} = R_{\alpha \beta} -\frac12 g_{\alpha \beta}\, R
=0$\cite{Inf57b}. Such interpretation, adopted by
Infeld\cite{Inf38}, permitted to simplify the entire deduction of
the equations of motion. The concrete mathematical model was
obtained in collaboration with Pleba\'nski\cite{Inf60,Inf56b}.

To fix the ideas, let $\xi^s$(t) be a world line and $\varphi$ a
scalar field that depends on coordinates $x^s$, time $x^0=t$, and
also on the $\xi^s(t)$ and their time derivatives
$\dot{\xi}^s(t)$: $\varphi = \varphi(x^s, t; \xi^s, \dot{\xi}^s)$.
As the procedure produces fields $\varphi$ which are singular on
the world lines $\xi^s(t)$, Infeld and Pleba\'nski looked for a
transformation theory changing $\varphi$ into a continuous
function $\widetilde \varphi$ of the $\xi^s$, $\dot{\xi}^s$ and
$\ddot{\xi}^s$, without recurse to the renormalization procedure.
Hence, they were faced with the problem of interpreting the
expression

\begin{equation}
\int \limits_{\Omega} \frac{\delta(\zeta)}{\vert \zeta \vert^k} \,
d\zeta, \quad k>0
\label{ipdelta}
\end{equation}
often considered as divergent. However, the diverse definitions of
the $\delta$--functions, as presented in the preceding
section\footnote{Infeld and Pleba\'nski categorize three different
methods: (A) axiomatic, essentially depicted by properties
(\ref{ddelta})--(\ref{dalg}); (B) Fourier transformation. This
lies on the definition (\ref{fourier}) and the preceding
properties; (C) The realistic method, which lies on the very core
of the Dirac's intuition (see the paragraphs between equations
(\ref{fourier}) and (\ref{gauss})) for which the axiomatic
properties (\ref{ddelta})--(\ref{dalg}) are immediately
justified\cite{Inf60,Inf56a,Inf57}.}, are useless to interpret
(\ref{ipdelta}). In all cases the integrands $f(\zeta)
\delta(\zeta)$ were considered for the continuous functions
$f(\zeta)$, at least in the vicinities of $\zeta =0$. Thus,
(\ref{ipdelta}) is simply meaningless! Infeld and Pleba\'nski
solved the problem by narrowing the definition of Dirac's
$\delta$--function so that the integral

\begin{equation}
\int \limits_{\Omega} \psi (\zeta) \, \Hat\delta_{IP} (\zeta) \,
d\zeta
\label{ipdelta2}
\end{equation}
acquires a definite meaning even if $\psi(\zeta)$ has a
singularity up to the $k$th order\cite{Inf56a,Inf57}. They
introduced their delta function $\Hat\delta_{IP}$ in an axiomatic
form which extends the limits of Schwarts distribution theory:

%%%%%%%%%%%%%%%%%%%%%%%%%%%%%%
\begin{itemize}
\item[]
($\mbox{\rm IP}_1$) $\Hat\delta_{IP}$ has all derivatives for
$\zeta \neq 0$.
%%%%%%%%%%%%%%%%%%%%%%%%%%%%%%
\item[]
($\mbox{\rm IP}_2$) $\Hat \delta_{IP}=0$ if $\zeta \neq 0$; $\Hat
\delta_{IP}(0)$ is undefined.
%%%%%%%%%%%%%%%%%%%%%%%%%%%%%%
\item[]
($\mbox{\rm IP}_3$) For a continuous function $f(\zeta)$:
\begin{equation}
f(0) = \int \limits_{\Omega} f(\zeta) \, \Hat\delta_{IP} (\zeta)
\, d\zeta.
\label{ipdelta2b}
\end{equation}
%%%%%%%%%%%%%%%%%%%%%%%%%%%%%%
\item[]
($\mbox{\rm IP}_4$) For a certain $k$ we have
\begin{equation}
\int \limits_{\Omega} \frac{\Hat\delta_{IP}(\zeta)}{\vert \zeta
\vert^k} \, d\zeta = \omega_k
\label{ipdelta3}
\end{equation}
where $\omega_k$ is a previously assigned value.
\end{itemize}
%%%%%%%%%%%%%%%%%%%%%%%%%%%%%%

\noindent The axioms ($\mbox{\rm IP}_1$)--($\mbox{\rm IP}_4$), all can
hold for a convenient class of functions $\Hat \delta_{IP}
(\varepsilon, \zeta)$ in the realistic approach\cite{Inf60}.
Hence, Infeld and Pleba\'nski defined the following algebraic
rules
\begin{equation}
\left\{
\begin{array}{l}
\Hat\delta_{IP}(\zeta) = (1 + \omega_k \, \vert \zeta \vert^k) \,
\delta_{IP} (\zeta)\\[4pt]
\delta_{IP} (\zeta) = \alpha \, \vert \zeta \vert^k \,
\frac{d}{d\zeta} (\zeta \, \delta (\zeta) )
\end{array}
\right.
\label{ipdelta4}
\end{equation}
where $\delta_{IP}$ corresponds to the choice $\omega_k=0$ in
(\ref{ipdelta3}), $\delta$ is the conventional Dirac function and
$\alpha$ is an infinite constant chosen such that\footnote{This is
certainly possible in the realistic method and $\alpha$ turns out
to be as singular as $\varepsilon^{-k}$. In their book, Infeld and
Pleba\'nski found $\alpha = \varepsilon^{-k} \sqrt{\pi} \, \left[
2^{(k+1)/2} \Gamma \left( \frac{k+1}{2} \right) \right]^{-1}$ for
a Gaussian $\delta_{IP} (\varepsilon, \zeta)$.}
\[
\int \limits_{\Omega} \delta_{IP} (\zeta) \, d\zeta = 1 = \int
\limits_{\Omega} \Hat\delta_{IP} (\zeta) \, d\zeta.
\]
The meaning of equations (\ref{ipdelta4}) is that their two sides
give equivalent results as factors in an integrand, exactly like
the ordinary Dirac's $\delta$--function. The
$\Hat\delta_{IP}$--functions, however, allow to associate definite
meanings with integrals of products of $\Hat \delta_{IP}$ with
functions that become divergent for $\zeta \rightarrow 0$. This is
why Infeld and Pleba\'nski called them ``good
functions''\cite{Inf60}. They claimed the application of these
functions as equivalent to the regularization procedure.

As for the transformation theory, Infeld and Pleba\'nski
generalized these {\it good functions\/} to more dimensions (which
is simple enough) and established\cite{Inf57b}
\[
\int \varphi \, \Hat \delta_{IP(3)} \, (x^s - \xi^s) \, d_{(3)} x
= \widetilde \varphi
\]
as the definition of $\widetilde \varphi$, where $\Hat
\delta_{IP(3)}$ is their three--dimensional good
$\delta$--function. Thus the tilde means two things: singularities
of $\varphi$ are ignored and, for $x^s$, the $\xi^s$ are
substituted.

%----------------------------------------------------------->ch

\section{Is the $\delta$--Zoology exhausted?}

It might seem that the $\delta$--Zoology (by paraphrasing
Heisenberg\cite{Meh01a}) is exhausted. Yet, from time to time,
distributions on differential domains are also considered. For
instance, a rigorous definition of the delta function could be
obtained, in the sense of Mikusi\'nski\cite{Mik48,Mik59}, by
defining the generalized functions as the closure of certain
ordinary functional spaces with respect to a weak topology. Last
years, the hyperfunctions of Sato\cite{Sat59} (considered more
general than the improper functions) have more and more
applications. In the present section, we shall analyze a set of
new objects which have been profitable in the context of Darboux
transformations in quantum mechanics (a topic which, under the
name of {\it factorization method\/}, was also investigated by
Dirac, Infeld, and Pleba\'nski). Let us start by remarking that
functions (\ref{gauss}) and (\ref{square}) behave as
$\varepsilon^{-1}$ for small values of $\varepsilon$. Now, what
about functions $\delta(\varepsilon, \zeta)$ showing an arbitrary
$\varepsilon$--dependence instead of $\varepsilon^{-1}$? Unlike
the Dirac's case, we shall take a family of ``well potentials''
\begin{equation}
\breve\delta(\varepsilon, \zeta) := \left\{
\begin{array}{ll}
-\varepsilon^{-2} \qquad & \vert \zeta \vert \leq \varepsilon\\[4pt]
0 \qquad & \vert \zeta \vert > \varepsilon.
\end{array}
\right. \label{nnro}
\end{equation}
In order to get a wider meaning of this new objects, let us
analyze
\begin{equation}
\Delta'(\varepsilon,\zeta):= \left\{
\begin{array}{ll}
\varepsilon^{-2} \qquad & \zeta \in (-\varepsilon, 0)\\[4pt]
-\varepsilon^{-2} \qquad & \zeta \in (0, \varepsilon)\\[4pt]
0 \qquad & \vert \zeta \vert > \varepsilon.
\end{array}
\right.
\label{derivative}
\end{equation}
which has been reported by the Christiansen's group\cite{Chr03}
and by Boykin\cite{Boy03} independently.  It is a matter of
integration (in the sense of distributions) to verify that
$\lim_{\varepsilon \rightarrow 0} \Delta'(\varepsilon,\zeta) =
\delta'(\zeta)$. Thus $\Delta'$ approaches the derivative of the
Dirac's delta function! The Christiansen's group worked on
definition (\ref{derivative}) to analyze the scattering properties
by regularizing finite--range potentials (point or contact
interactions). Their approach leads to the conclusion that
$\delta'(\zeta)$ is a transparent potential as opposite to the
Seba's theorem\cite{Seb86} which establishes that $\delta'(\zeta)$
should have zero transparency. They also studied second and third
order derivatives of the delta function. On the other hand, Boykin
obtained (\ref{derivative}) by conveniently transforming a finite
difference formula. He used a three dimension version of $\Delta'$
to derive the Gauss' law in a dielectric medium directly from the
charge densities, without using potentials\cite{Boy03}.

Now, as we can see, our well potentials (\ref{nnro}) resemble
those in (\ref{derivative}). In some sense, we could interpret
either the limit $\breve \delta(\varepsilon,\zeta)_{\varepsilon
\rightarrow 0}$ as an ``incomplete'' derivative of the delta
function or $\Delta'(\varepsilon, \zeta)$ as a combination of
$\breve \delta(\varepsilon, \zeta)$:
\[
\Delta'(\varepsilon, \zeta) = \breve \delta (\varepsilon, \zeta)
\, \Theta(\zeta) - \breve \delta (\varepsilon, \zeta) \, \Theta
(-\zeta).
\]
Remark that, for dealing with test functions $f$, the sequence
$\breve \delta (\varepsilon, \zeta)$ guarantees small domains of
integration $\Omega$, centered at the origin\footnote{Indeed,
$\Omega$ does not need to be centered at the origin but at an
arbitrary accumulation point $\zeta_0$. I prefer to use
$\zeta_0=0$.}. Henceforth, consider the following result

\begin{equation}
\int \limits_{\Omega} \, f(\zeta) \, \vert \zeta \vert \, \breve
\delta (\varepsilon, \zeta) \, d\zeta = -f(\xi)
\label{orodelta}
\end{equation}
where $\xi \in (-\varepsilon, \varepsilon)$, the function
$f(\zeta)$ is differentiable enough and the mean value theorem for
integration has been applied. By calculating the limit
$\varepsilon \rightarrow 0$ and interchanging the limiting process
with integration in (\ref{orodelta}), one establishes the
following rule of manipulating the $\breve\delta$--function:
\begin{equation}
\breve \delta (\zeta) \, \vert \zeta \vert= - \delta(\zeta)
\label{oroalg}
\end{equation}
with $\delta$ the ordinary Dirac's function. Now, let us draw our
attention to the first of equations (\ref{dalg}). It shows that
{\em whenever one divides both sides of an equation by a variable
$\zeta$ which can take on the value zero, one should add on to one
side an arbitrary multiple of $\delta(\zeta)$} (see
Dirac\cite{Dir58}), just as it occurs for the derivative of the
$\log(\zeta)$ function: $d\log(\zeta)/d\zeta=1/\zeta - i\pi \,
\delta(\zeta)$. Thus, equation (\ref{oroalg}) becomes:
\begin{equation}
\breve \delta (\zeta) = -\frac{\delta(\zeta)}{\vert \zeta \vert} +
c \, \delta(\zeta) \, \mbox{\rm sgn}(\zeta)
\label{oroalg2}
\end{equation}
where $c$ is an arbitrary constant and $\mbox{\rm sgn}(\zeta)$ is
the {\it sign\/} improper function. The right hand side of
(\ref{oroalg2}) is neither a definition of $\breve \delta(\zeta)$
nor an equality {\it sensu stricto\/}. As before, equations
(\ref{oroalg}) and (\ref{oroalg2}) are merely operational
equivalences requiring the integration. Nonetheless,
$\breve\delta$ has a stronger divergence than $\delta$ at
$\zeta=0$.

What can we accomplish with these new $\breve\delta$ `improper
functions'? Let us take a continuous function $\phi(\zeta)$ which
is not necessarily differentiable at $\zeta=0$ but such that
$\phi(\zeta) \sim const \vert \zeta \vert$ for $\zeta \rightarrow
0$. Then, the following transformation holds:
\begin{equation}
\int \limits_{\Omega} \phi(\zeta) \, \breve\delta(\zeta) \, d\zeta
= \breve\phi(0)
\label{cuerno}
\end{equation}
where $\breve \phi (\xi)$ is a new continuous and differentiable
function in all the real line (remember we have taken the
accumulation point $\zeta_0$ equal to zero). As an immediate
example one can substitute $\phi (\zeta)$ for $f(\zeta) \, \vert
\zeta \vert$ and $\breve\phi(\xi)$ for $-f(\xi)$ in equation
(\ref{orodelta}); after the usual limit procedure one gets
(\ref{cuerno}).

Observe that, in the previous derivations, the class of functions
$\breve \delta (\varepsilon,\zeta)$ have been considered on a {\it
free particle background\/}. The relevance of the singular $\breve
\delta$ `function' is analogous on a nontrivial background. In
particular, let $V(\zeta)$ be a singular, one dimensional,
potential growing as $\zeta^{-2}$ for $\zeta \rightarrow 0$. The
new potential $V_{reg}(\varepsilon,\zeta) = V(\zeta) + \alpha
\breve \delta(\varepsilon,\zeta)$ can be proved to be regular at
$\zeta = 0$ for the appropriate value of the strength $\alpha$ and
any value of $\varepsilon \neq 0$. Recent results show that
periodic singular potentials admit a regularization procedure in
this sense\cite{Neg03}. If the initial potential is the Scarf's
one: $V^{s}(\zeta)=V_0/\sin^2(\zeta)$, then
$V^s_{reg}(\varepsilon,\zeta)$ is a family of regular potentials
such that $\lim_{\varepsilon \rightarrow 0}
V^s_{reg}(\varepsilon,\zeta) = V^s(\zeta)+ \breve \delta(\zeta)$.
Similar results are obtained for other singular potentials defined
on the complete real line, including the cases of discrete
spectrum (see e.g. Dutt et al\cite{Dut01}, and Negro et
al\cite{Negp}). Furthermore, it has been shown that this procedure
does not change the results of a supersymmetric transformation.
Thus, the Darboux transformations and the $\breve
\delta$--regularization procedure commute in quantum
mechanics\cite{Neg03,Negp}.

Finally, let us remark on the fact that generalized functions can
be represented as sequences of ordinary functions which converge
in a certain way. This property, as we have seen, stimulated the
development of the theory of distributions and related approaches.
The use of improper functions thus expands the range of problems
that can be tackled in mathematical and theoretical physics, in
particular in the theory of differential equations and quantum
physics.

%----------------------------------------------------------->Ack

\section*{Acknowledgements}

I owe the manuscript's typing to Miriam Lomeli. I am grateful to
the Organizers of the Pleba\'nski's Festival for their kind
invitation to give a talk in the Conference, specially to Maciej
Przanowski and Bogdan Mielnik, my thanks. This work has been
supported by CONACyT (Mexico) and MCYT (Spain).

%----------------------------------------------------------->bib

\end{document}